\documentstyle[12pt]{article}

\begin{document}

\vspace{1cm}
{\large
Charge Fluctuations and Magnetic Inelastic Neutron Scattering in Copper-Oxide high -Tc  superconductors.
}

\vspace{0.5cm}
{\it A. S. Moskvin, A. S. Ovchinnikov  and O. S. Kovalev.}

\vspace{0.3cm}
{\small
Department of Theoretical Physics, Ural State University,
620083, Lenin Ave. 51, Ekaterinburg , Russia.
}
\vspace{0.5cm}

{\small
Spin subsystem  of the copper oxides within the polar Jahn-Teller centers model corresponds to a singlet-triplet magnet where the local boson movement accompanied by the induced longitudinal spin fluctuations. These fluctuations determine the  features of the magnetic inelastic scattering in the high-Tc cuprates. 
}
\vspace{1cm}

The $CuO_{4}$-cluster based copper oxides are considered as a system unstable with respect to the disproportionation with the formation of the polar Jahn-Teller (JT) centers phase being a system of local bosons (coupled electron singlet pairs) moving in a lattice of singlet-triplet $[CuO4]^{5}$-JT-centers. These centers are characterized by  a near degeneracy  within the ground state manifold  involving  $^{1}{A}_{1g}$ , $^{1,3}{E}_{u}$ terms distinguished by the spin multiplicity, parity and orbital degeneracy that provides unconventional behaviour for the hole centers $CuO_{4}^{5-}$ with an active interplay of various modes. Generally, we deal with complicated quasi-two-dimensional multi-component strongly correlated quantum bose-liquid.
The spin subsystem in the JT-centers phase is the generalized singlet-triplet magnetic with the induced spin fluctuations that unlike the conventional spin-orientational (transversal) fluctuations are connected with fluctuations of spin density or spin multiplicity (longitudinal spin fluctuations). The main source of the spin multiplicity fluctuations is determined by a dependence of the singlet-triplet separation $\Delta_{ST} = \varepsilon (3Eu) - \varepsilon (1A1g) \sim 0.1 eV$ on the local boson density $\Delta_{st}(m) =  \Delta_{st} + \sum_{m} \Delta(mn) \delta N(n)$, where $\delta N(n) = N(n) - \langle N(n) \rangle $ is a charge fluctuation.
In general, the coupling between the induced spin fluctuations and the charge fluctuations can be represented in terms of the corresponding correlations functions as
$$
	{\langle S_{z}(\vec q) S_{z}(- \vec q)\rangle}^{ind}_{\omega} =
\int F(\vec q, \vec Q)G(\vec Q) {\langle \delta N(\vec Q) \delta N(-\vec Q)\rangle}_{\omega} d {\vec Q}
$$
  where $F(\vec q, \vec Q) \sim {\langle S_{z} (\vec q - \vec Q) S_{z} (\vec Q - \vec q)\rangle}_{0}$  is an effective spin form-factor determined by the longitudinal static spin correlation function calculated within an averaged charge distribution, $G(\vec Q)$ is a crystalline field structural factor. 

This integral relation permits to explain many unusual spectral and momentum $(\omega, \vec q)$, temperature and concentration $(x,T)$ dependencies for different effects determined by spin correlations functions including inelastic neutron scattering (INS), spin-lattice relaxation, static spin susceptibility. 
A charge subsystem in the copper oxides is coupled not only with the spin and local JT-modes but with the conventional atomic displacements modes determining the induced structural fluctuations:  ${\delta {\vec u}(i)}= {\sum}_{j} {\vec B}_{ij} \delta N(j)$.
A wide number of the experimentally observed peculariaties is connected with the effect of the induced displacement including an observation of the dynamic and static incommensurate superstructures (stripe-structures) in $YBa_{2}Cu_{3}O_{6.93}$ [1], $La_{2-x}Sr_{x}CuO_{4}$ [2], $(LaNd)_{2-x}Sr_{x}CuO_{4}$ [2], directly connected with corresponding charge (boson) modes. With taking account of $xy$-symmetry in the $CuO_{2}$-planes the resultant stripe superstructure will 
be described by the  $\vec q$-vectors: $(±2 \varepsilon, 0)$ and $(0, ±2 \varepsilon)$. The stripe structured charge fluctuations can 
result in the induced longitudinal spin fluctuations with $\vec q$-dependence «peaked» at  $ {\vec q} =((\frac{1}{2} ± \varepsilon), \frac{1}{2})$, $( \frac{1}{2}, (\frac{1}{2} ± \varepsilon))$ for the antiferromagnetic spin-spin interactions. Fig. 1 gives an example of the «four-peaks» $(q_{x}, q_{y})$ - dependence of the induced spin fluctuations for the copper oxide like $La_{2-x}Sr_{x}CuO_{4}$ calculated within a gaussian approximation for the spin and charge correlation functions.

With taking account of the phase separation the INS spectra are treated in general within the "three-oscillators" model with i) spin contribution of parent dielectric phase, ii) charge fluctuations of polar center phase  and iii) that of phonon oscillator. In the result of the fitting the corresponding frequencies, the relaxation times and the effective correlation lengths, the typical sizes of  micro-regions of dielectric phase, their concentration and  temperature dependencies were found. As an illustration in Fig.2, 3 we presented the examples of the model fitting of the experimental INS data  for 1-2-3 system. The INS spectral dependence for the underdoped system $(x=6.51)$ shows a competition of the spin (dashed) and charge (dotted) oscillators contributions as a 
result of the phase separation. The optimally doped system $(x=6.92)$ does not reveal the parent phase contribution and the INS intensity reflects a result of the interaction between the charge and the 41 meV -phonon modes (a polariton effect). Fig. 4 presents temperature dependence of the charge oscillator frequency $w_{ch}$ for different 123-systems.
Along with the close  $w_{ch}$  values note the frequency shift and the frequency splitting effect (see inset in Fig. 4) obtained as a result of the fitting. Dynamic charge fluctuations observed in the 123-system [3] at $\hbar \omega ~ 32 meV$ due to the induced spin fluctuation effect are naturally coupled with a charge (boson) transfer between nearest $CuO_{2}$-planes. An induced spin fluctuations effect can be used for studying the charge system (or that of local bosons) with the help of the magnetic methods. As a whole a spin subsystem in the HTSC's reveals very unusual properties but nevertheless we are firmly convinced it does not play a principal role in determining the origin of the superconductivity itself.

\vspace{2cm}
REFERENCES
\vspace{0.5cm}

[1] H.A.Mook et al., Phys. Rev. Lett. {\bf 77}, 370 (1996).

[2] J.M.Tranquada et al., Phys.Rev.B {\bf 54}, 7489 (1996).

[3] F. Onufrieva and J. Rossat-Mignod. Phys.Rev.B {\bf 52}, 7572 (1995).

\end{document}